\documentclass[12pt]{iopart}

\usepackage{iopams}
\usepackage{graphicx}
\usepackage{stmaryrd}

\newcommand{\id}{\textrm{d}}

\newcommand{\bbI}{{\mathbb I}}
\newcommand{\bbG}{{\mathbb G}}
\newcommand{\bbR}{{\mathbb R}}

\newcommand{\caI}{{\mathcal I}}
\newcommand{\caL}{{\mathcal L}}
\newcommand{\caO}{{\mathcal O}}
\newcommand{\caS}{{\mathcal S}}
\newcommand{\caT}{{\mathcal T}}
\newcommand{\caU}{{\mathcal U}}

\newcommand{\mean}[1]{{\left< #1 \right>}}
\newcommand{\pfrac}[2]{\left[\frac{#1}{#2}\right]}

\begin{document}

\title{An update on nonequilibrium linear response}

\author{M Baiesi$^{1,2}$ and C Maes$^3$}
\address{$^1$ Dipartimento di Fisica ed Astronomia, Universit\`a di Padova, Via Marzolo 8, I-35131 Padova, Italy}
\address{$^2$ INFN, Sezione di Padova, Via Marzolo 8, I-35131 Padova, Italy}
\ead{baiesi@pd.infn.it}

\address{$^3$ Instituut voor Theoretische Fysica, K.U. Leuven, Celestijnenlaad 400D, Leuven, Belgium}
\ead{christian.maes@fys.kuleuven.be}

\begin{abstract}
The unique fluctuation-dissipation theorem for equilibrium stands in  contrast
with the wide variety of nonequilibrium linear response formul{\ae}.  
Their most traditional approach is ``analytic'', which, in the absence of detailed balance,
introduces the logarithm of the stationary probability density as observable.  
The theory of dynamical systems offers an alternative with a formula 
that continues to work when the stationary distribution is not smooth.
We show that this method works equally well for stochastic dynamics, and 
we illustrate it with a numerical example for the perturbation of circadian cycles. 
A second ``probabilistic'' approach starts from dynamical ensembles
and expands the probability weights on path space. This line suggests new
physical questions, as we meet the {\it frenetic} 
contribution to linear response, and the relevance of the change in dynamical 
activity in the relaxation to a (new) nonequilibrium condition.
\end{abstract}

\pacs{05.70.Ln,  
05.40.-a,        
05.20.-y         
}
\maketitle

\section{Introduction}\label{intro}

An open system in contact with a large environment is in stationary equilibrium for a given reduced scale of description when at each moment its condition realizes minimal free energy. The total entropy production is then zero and the evolution is time-reversal invariant.   Small disturbances break the stationary behavior and after some time an eventually new equilibrium is established mainly via dissipative effects.  The comparison with the original equilibrium is the domain of linear response: since about 60 years now~\cite{cal51,kub66}, 
for classical, quantum, open, and closed systems basically the same formula relates the
response to a small perturbation with an equilibrium correlation function. 
This formula (later called Kubo formula) exactly picks up the physical interpretation 
in terms of dissipation, hence the name fluctuation-dissipation theorem~\cite{tod92}.

Response theory continues making sense for stimuli to nonequilibrium systems, where entropy is being produced already before the perturbation.   Over the years various types of linear response formul{\ae} have indeed been obtained for nonequilibria, including rather diverse  but specific
models such as in climatology and for glassy or coarsening dynamics.  For better orientation the present paper identifies some common traits between the various approaches to put them in a more unified framework.  Suppose for example that one wants a numerical code for predicting the response of an
out-of-equilibrium systems, i.e., without actually perturbing the system.
Without a clear picture of the available theoretical results, the choice of the method, if any, would be limited and possibly suboptimal. To answer requests like these, we aim at classifying the features of several ``extended fluctuation-dissipation theorems'' for nonequilibrium, highlight briefly their strengths and weaknesses, and the eventual relations between them.

Reviews and classifications of nonequilibrium response already exist. The paper by Marini Bettolo Marconi, Puglisi, Rondoni, and Vulpiani has reviewed response in statistical physics in the light of recent fluctuation relations~\cite{mar08}. Seifert and Speck have introduced a classification of some fluctuation-dissipation theorems into three classes~\cite{sei10}.  Chetrite and Gupta present a more mathematical view~\cite{che11}.
The present work includes the case of deterministic dynamics and attempts a concise classification of linear response for nonequilibrium.
We will conclude that there are three main classes.  The first class of formul{\ae} can be derived from the Kubo-Agarwal formula, which itself starts from a Dyson-expansion of the perturbed Markov semi-group.  A second approach originates in the theory of dynamical systems and can be applied when the stationary distribution is not smooth; it gives rise to a numerical algorithm which in fact also applies for certain
stochastic dynamics, as we show.  The third class is much more probabilistic and treats noise as an important observable in the linear response. The synthesis is there provided by introducing the excess in dynamical activity in a second ``frenetic'' contribution to the traditional Kubo fluctuation-dissipation formula.  In that sense, the name fluctuation--{\it dissipation} theorem (even ``extended'') is not fully suitable
for nonequilibrium systems, as e.g.\ their return to stationary nonequilibrium is not uniquely characterized by dissipation~\cite{mae11}.  On the mathematical side, our classification is to work either with generators (for Markov evolutions) or to work with weights on path space (mostly still limited to classical dynamical ensembles).  Referring to those, we call it the analytic approach (Section \ref{per}) versus the probabilistic approach 
(Section \ref{pro}). Remarks complementing those in previous sections, and conclusions, 
are found in Section \ref{rema}.

\section{Analytic approach}\label{per}
The framework we consider is that of Markov dynamics (no memory) on regions of $\bbR^d$. 
That includes deterministic dynamical systems, being essentially first order in time.
It also includes jump processes as described with Master equations, 
but we use here the language of diffusion because it reduces naturally to that
of deterministic dynamical systems in the limit of zero noise, realized
here with a state-independent diffusion constant $D \to 0$.  
For even much greater simplicity we choose overdamped diffusions where the velocity field
\begin{eqnarray}
\label{lan}
\dot{x}_t = &v(x_t)& \quad\mbox{is given by}\\
             &v(x_t)&  = F(x_t) + \sqrt{2 D }\, \xi_t\nonumber
\end{eqnarray}
for standard white noise $\xi_t$~\cite{ris89}.  In Section  \ref{path} we also treat (underdamped or inertial) 
Langevin processes. In all these cases it makes sense to speak about the so called backward generator $L$, 
working on observables (Heisenberg picture).
The expectation of an observable $A$ at time $t\geq 0$ is then given by
\begin{equation}\label{gena}
 \mean{A(x_t)} = \int \rho(dx)  \left( e^{t L} A\right)(x)
\end{equation}
when at time zero the states are distributed with probability $\rho$, possibly singular with respect to the reference volume element $dx$ on state space.  We can also abbreviate that as $\id\mean{A(t)} / \id t = \mean{LA\,(t)}$.
When no confusion can arise, we continue to write $A(t) =A(x_t)$ for the (most often random) value of the observable at time $t\geq 0$, and $\langle B(0) A(t)\rangle$ for the time-correlation between observable $B$ at time zero and observable $A$ at time $t$.

For the overdamped diffusion (\ref{lan}) the generator is
\begin{equation}\label{gene}
(LA)\,(x) = (F \cdot \nabla A)\,(x) + D \,\Delta A\,(x)
\end{equation}
A perturbation changes the drift $F \to F^h\equiv F+h F_1$ where,
for simplicity, we avoid inserting a time-dependence in the small amplitude $h$,
applied at all times $t > 0$. This leads to a perturbed backward generator $L^h \equiv L + h L_1$ with
$L_1 A= F_1 \cdot \nabla A$.
The change in expectations at times $t$ with respect to what we had for the unperturbed dynamics follows 
from \eref{gena}:
 \begin{equation}
\langle A(t)\rangle^h - \langle A(t)\rangle = \int \rho(\id x)\,
\,\big(e^{tL^h} - e^{tL}\big)A\,(x)
 \end{equation}

To linear order in $h$,
 \begin{equation}
e^{tL^h} - e^{tL} = \int_0^t e^{sL}\,(L^h-L)\,e^{(t-s)L} \id s + \caO(h^2)
 \end{equation}
yielding a linear response
\begin{eqnarray}
\frac{\id \langle A(t)\rangle^h}{\id h}{\bigg|}_{h=0}
 &\equiv& \lim_{h\to 0} \frac{\langle A(t)\rangle^h - \langle A(t)\rangle}{h} \nonumber\\
&\equiv& 
\chi(t) = \int_0^t \id s \,R(t,s)
\end{eqnarray}
including a susceptibility $\chi(t)$ as the integration of a response function of the form
\begin{equation}
\label{dys}
\fbox {$ \displaystyle
R(t,s) = \int \rho(\id x)\;\big(e^{sL}\,L_1\,e^{(t-s)L}\,A\big)\,(x)
$}
\end{equation}
More generally, when applying a time-dependent perturbation $h_s$ at time $s>0$ we also have
 \begin{equation}
\frac{\delta \langle A(t)\rangle^h}{\delta h_s}{\bigg|}_{h=0} = R(t,s),\quad s<t
 \end{equation}
We mostly restrict ourselves to the case where $\rho$ is stationary:
the response depends only on the time difference $\tau=t-s$ and we can write
\begin{equation}
\fbox {$ \displaystyle
R(\tau) = \int\, \rho(\id x)\,L_1 e^{\tau L}A\,(x) \label{dys2a}
$}
\end{equation}
as the central object of study for the linear response of stationary Markov evolutions within an analytic approach\footnote{
We can of course also use (\ref{dys2a}) for discrete jump processes. For each pair $x,y$ of states,
the perturbation enters as a modification of jump rates $w(x,y)\to w(x,y)+h\,w_1(x,y)$.  
The matrix $L_1$ to be put in (\ref{dys2a}) has elements $w_1(x,y)$ if $x\ne y$, and $w_1(x,x)=-\sum_y w_1(x,y)$.}.

 The direct reading of the right-hand side of \eref{dys2a}, further discussed under Section \ref{sec:det},
is that $e^{\tau L}A$ evolves the observable $A$ for a time $\tau$ and then $L_1$ acts on the result to evaluate it in state $x$.
However, we start in the next section with the more frequent
approach of acting on $\rho$.

\subsection{Acting on probabilities}\label{apr}
In this section we focus on manipulations with the stationary probability distribution $\rho$. The basic step from \eref{dys2a} is partial integration, which means that it is assumed here that $\rho$ has a smooth density with respect to the reference volume element, $\rho(\id x) = \rho(x) \,\id x$.  In many cases that appears to be  a reasonable physical assumption when the level of description is mesoscopic to macroscopic, independent of whether the system is driven or not (see \cite{col12}).

\subsubsection{Kubo--Agarwal formula}
Assuming a smooth density $\rho$ we have that \eref{dys2a} can be rewritten as
\begin{equation}
\label{agar0}
R(\tau) = \int\, \id x\, \rho(x)\,\frac{\caL_1\rho}{\rho}(x)\, e^{\tau L}A\,(x)
\end{equation}
where the adjoint $\caL_1$ is defined by
$\int \id x (\caL_1\rho)(x)\,A(x) = \int \id x \rho(x)\,(L_1 A)(x)$.
 Adjoints are forward generators of the time-evolution on densities,
 as appears e.g. in Master equations. 
For the diffusion \eref{gene} the adjoint of $L$ is the operator of the Fokker-Planck equation 
($\partial_t\rho = \caL \rho$, see~\cite{ris89})
\begin{equation}\label{FPL}
{\cal L} \rho = -\nabla\cdot(F\rho)+\Delta(D \rho)
\end{equation} 
and ${\cal L}_1 \rho = -\nabla\cdot(F_1\rho)$ so that
 \eref{agar0} takes the form
\begin{equation}
\label{difag}
R(\tau) = -\mean{ [\nabla \cdot F_1(0) + F_1(0)\cdot\nabla \log \rho(0)]\,A(\tau)}
\end{equation} 
which is a specific realization of 
  \begin{equation}\label{agar}
\fbox {$ \displaystyle R(\tau) = \mean{ \frac{\caL_1\rho}{\rho}(0)\,A(\tau)} $}
= \langle B(0)\,A(\tau)\rangle
\end{equation}
with {\it observable} $B(x) = \frac{\caL_1\rho}{\rho}(x)$.  Note that in general the stationary expectation $\langle B\rangle =0$ because $\int \id x\, \caL_1 \rho(x) = 0$ from the normalization of $\rho$.
Applications of that Agarwal formula~\cite{aga72} in practice meet the difficulty of needing to know the density $\rho$ (which is usually not available) and the details of the dynamics for $\caL_1$. It is thus a result (from partial integration) on a formal level.

Formula \eref{agar} is associated to equilibrium, see formula (2.5) in~\cite{rue09}.  
Of course we have only used that $\rho$ is smooth.  It is easy to verify that
in the case of detailed balance with conservative forces $F=-\nabla U$ at temperature $D=1/\beta$ we obtain the Kubo formula \cite{kub66} for the linear response relation in equilibrium.  Indeed, say for \eref{gene} with perturbation $F_1=-\nabla V$ around equilibrium $\rho \propto \exp[-\beta U]$, we get
 \begin{equation}
\frac{{\cal L}_1 \rho}{\rho} = -\beta \nabla V\cdot \nabla U + \Delta V = \beta\, LV
 \end{equation}
so that
\begin{equation}\label{kubo}
\mean{ \frac{\caL_1\rho}{\rho}(0)\,A(\tau)} = \beta\,\langle LV(0)\,A(\tau)\rangle = \beta\,\frac{\id}{\id \tau}\langle V(0)\, A(\tau) \rangle
\end{equation}
which is the (classical) Kubo formula.  The last equality has used that under detailed balance $\langle LV(0)\,A(\tau)\rangle =\langle V(0)\,LA(\tau)\rangle$ (see further details in Sec.~\ref{velo}).  In a way, the Agarwal formula \eref{agar} repeats Kubo's original derivation while stopping short before specifying $\rho$.

 Also others have re-found the Agarwal formula, such as in Theorem 2, formul{\ae} (2.22) and (2.23), of Chapter 2 in \cite{maj05}.  Weidlich gives a quantum version: his equation (2.17) replaces $\caL_1 \rho \rightarrow \frac{i}{\hbar}[\rho,H_1]$ with the commutator of the perturbing Hamiltonian $H_1$~\cite{wei71}.
H\"anggi and Thomas in \cite{han82} find the Agarwal formula in their equation (3.12) for time-dependent processes.
In the review~\cite{mar08} formula (2.70)--(2.73) is the Agarwal formula.  We also find these formul{\ae} such as \eref{difag} involving $\log \rho$ presented in the book by Risken, formula (7.10)--(7.13) in \cite{ris89}, and as formula (7) in \cite{spe06}.  There also re-started the emphasis on $\log \rho$ as ``generalized'' potential. 

\subsubsection{Information potential}
The formula that Falcioni, Isola and Vulpiani~\cite{fal90} derived for tiny displacements of the initial
condition along a unit vector $\hat{e}$,
\begin{equation}\label{vul}
\fbox {$ \displaystyle R(\tau) = \mean{ \hat{e}\cdot\nabla\log \rho (0)\,A(\tau)} $}
\end{equation}
(see it also as formula (3.13) in the review \cite{mar08})
corresponds to the case of diffusion with constant perturbation $F_1=-\hat{e}$ (impulsive constant force in the direction 
$-\hat{e}$) in \eref{difag}. Again, as there is no explicit dependence on the noise level the formula can be readily tried for chaotic dynamics with smooth stationary density~\cite{fal95}.  The function $\caI \equiv -\log \rho$ is sometimes called the information potential.  In fact, that potential gets a prominent place in various works on nonequilibrium linear response that follow the Hatano-Sasa formalism \cite{hat01}, such as in the more recent \cite{ver11} or \cite{pro09}.

Other formul{\ae} focusing on $\caI $ can be found in the works by Prost, Joanny and Parrondo~\cite{pro09}
and by Speck and Seifert~\cite{sei10}.
Consider the stationary density $\rho^h$ for the perturbed dynamics with generator $L^h$ 
and the linear response $\rho^h(x) - \rho(x) = h\,\rho_1(x) + \caO(h^2)$. 
From stationarity $\caL^h\rho^h =0$ we get $\caL_1\rho + \caL\rho_1=0$.
For the Agarwal formula \eref{agar} we need
 \begin{equation}
B = \frac{\caL_1\rho}{\rho} =  -\frac{\caL\rho_1}{\rho}
 \end{equation}
On the other hand,
 \begin{equation}
-\partial_h \caI^h = \partial_h \log \rho^h = \lim_{h\to 0}\frac{\rho^h - \rho}{h\,\rho^h} = \frac{\rho_1}{\rho}
 \end{equation}
($\partial_h$ is understood in $h=0$) so that for inserting in \eref{agar}, 
$B= \caL(\rho\,\partial_h \caI^h ) / \rho$. We conclude
 \begin{eqnarray}
R(\tau) &=& \int \id x  \caL(\rho\,\partial_h \caI^h )(x) e^{\tau L}A (x)\nonumber\\
&=& \int \id x  \,\rho(x)\,\partial_h \caI^h(x)\,Le^{\tau L}A (x) \nonumber\\
&=& \frac{\id}{\id \tau}\int \id x  \,\rho(x)\,\partial_h \caI^h(x)\,e^{\tau L}A (x)
 \end{eqnarray}
or
\begin{equation}\label{sei}
\fbox {$ \displaystyle
R(\tau) = \frac{\id}{\id \tau}\langle \partial_h {\caI}^h(0)\,A(\tau)\rangle
$}
\end{equation}
This formula with the special emphasis on the presence of the information potential appears in
\cite{sei10} (but with a different time-derivative), where $\caI$ is called stochastic entropy.  The result of \cite{pro09} starts from expanding the Hatano-Sasa identity, which effectively makes a special choice for the observable $A$.  There one imagines the process and hence also the $\cal I$ to depend on a family of parameters $\lambda$ and one asks how the expected value of $A= \partial_\lambda \cal I$ at $\lambda=\lambda^*$ changes under a small change in these parameters around a given $\lambda^*$.  The answer is provided by \eref{sei}, which is equation (5) in \cite{pro09}.

As already noticed for (\ref{agar}) 
a disadvantage of incorporating the information potential into the correlation function
 of \eref{vul} and \eref{sei} is that it is generally unknown and not directly measurable. 
Not only do we not know the stationary density, but also no clear physically relevant interpretation 
in terms of heat, work or thermodynamic potential has been found for $\caI$.
On the plus side, a clear advantage of (\ref{vul}) and \eref{sei} 
is that no detailed information on the dynamics is needed and that one can use parametrized forms of the stationary density $\rho$ to put in $\cal I$. 
Typically, (quasi-)Gaussian approximations are tried for $\rho$ 
to make formul{\ae} \eref{difag}, \eref{vul}, \eref{sei} more explicit and practically useful; 
see e.g. \cite{lei75,maj05,abr07,lac09}.  Note that there is no reference in formula \eref{difag} to the noise strength except through the stationary $\rho$.  If one replaces the $\rho$ in the $B$ of \eref{agar} by a Gaussian, its variance will effectively reflect the noise level but needs to be fitted. 

Finally, recent works on time-dependent processes with feedback rewrite the positivity of the entropy production in terms of the expected information potential (relative entropies) which appears useful for understanding work relations, see e.g. \cite{sag09,toy10}.
At any rate, \eref{sei} is the climax of the analytic approach for smooth probability densities, preserving the Kubo form \eref{kubo} most faithfully.

\subsubsection{State velocity}
\label{velo}
There is a relation between the information potential $\caI$
and the state-space velocity $u$.  
For diffusion processes with generator \eref{gene} this state velocity is
\begin{equation}\label{pv}
u \equiv \frac{j_\rho}{\rho} = F - D\nabla \log \rho = F + D\nabla \caI
\end{equation}
as can be readily checked from the expression for the stationary probability current $j_\rho$ appearing in the Smoluchowski equation $\partial_t \mu_t + \nabla j_{\mu_t} =0$ expressing conservation of probability $\mu_t$.
Of course this probability velocity needs not be related to physical currents, but an interesting observation writes a nonequilibrium response formula as a co-moving (equilibrium) fluctuation-dissipation theorem.

In a stationary process
 two quantities $A$ and $B$ have time-translationally invariant
correlations $\mean{B(0)A(t)}=\mean{B(-t)A(0)}$.  As we mentioned above, the Kubo formula (or, the fluctuation-dissipation theorem) holds under detailed balance, i.e., for an unperturbed evolution which gives rise not only to a stationary but also to a time-symmetric distribution of trajectories.  Nonequilibrium means breaking time-reversal invariance.  We introduce the operator $L^*$ that generates the time-reversed motion to describe for example
 \begin{equation}
\mean{B(-t)A(0)} = \int  \rho(\id x)\, A(x) \left(e^{t L^*}B\right)(x),\quad t\geq 0
 \end{equation}
One can alternatively write
$\int \rho(\id x)\,(Lf)(x)\,g(x) = \int  \rho(\id x)\, f(x)\,(L^*g)(x)$, 
from which we see that $L^*g = \caL(\rho g)/\rho$.  
For the overdamped diffusion processes that we have considered so far, from \eref{gene} and \eref{FPL},
$L^*= L -2u\cdot \nabla = -F\nabla + D \Delta  + 2  D \nabla \log \rho \nabla $.

Note now that the Kubo formula \eref{kubo} would follow from the Agarwal formula 
when in \eref{agar} we  take $B = -\beta L^*V$.  Indeed,
 \begin{equation}
\langle L^*V(0) A(\tau)\rangle = 
\langle V(0) L A(\tau)\rangle = \frac{\id}{\id \tau}\langle V(0) A(\tau)\rangle 
 \end{equation}
Detailed balance corresponds to $L=L^*$ (i.e., $\mean{B(-t)A(0)} = \mean{B(t)A(0)}$) and
corrections to the Kubo formula will thus arise from choosing in \eref{agar}
\begin{equation}\label{loc}
B(x) = -\frac{L+L^*}{2}V\,(x) = -L^*V(x) -  \frac{L-L^*}{2}V\,(x)
\end{equation}
for some function $V$.
In other words the antisymmetric part $\frac{L-L^*}{2}V$ in the formula \eref{loc} for $B$ will be responsible for violating the Kubo formula and will show the nonequilibrium aspect.  Moreover, $L- L^* = 2 u\cdot\nabla$ relates to the state velocity \eref{pv}.  Substituting these relations via \eref{loc} in \eref{agar}, 
we find that for diffusions \eref{gene} (with $D=1/\beta$)
\begin{eqnarray}\label{fra1}
R(t,s) &=& \beta \frac{\id}{\id s}\int \id x\, \rho(x)\,
V(x)\, e^{(t-s)L}A(x)\nonumber\\
&&  -\beta\int \id x\, \rho(x)\, u(x) \cdot \nabla V(x)\, e^{(t-s)L}A(x)
\end{eqnarray}
abbreviated as
\begin{equation}\label{fra}
\fbox {$ \displaystyle
R(t,s) = \beta\frac{\id}{\id s}\mean{V(s)A(t)}  -
\beta \mean{ [u(s) \cdot\nabla V(s)] A(t)}
$}
\end{equation}
The equation \eref{fra} shows that the equilibrium Kubo form gets ``restored'' when
describing the system in the Lagrangian frame moving with drift
velocity $u$. The
passage to the Lagrangian frame of local velocity $\id/\id s \rightarrow \id/\id s - u\cdot \nabla$ ``removes'' the
non-conservative forcing from the formula, as explained in \cite{spe06,che08,che09}.
Still, if we do not know the stationary $\rho$, the formula \eref{fra} implies a statistical average over the unknown
vector $u$. 

\subsection{Acting on the observable}
\label{sec:det}
We go back to the original \eref{dys2a} and we move to an algorithm focusing on the evolution of the observable $A$ rather than that of the density $\rho$.  Various contributions to the theory of dynamical systems start exactly from that formula for the formulation of linear response as there still no assumption is needed on the smoothness of the probability $\rho$. 

The main player now is $L_1 e^{tL} A$ in (\ref{dys}), and $L$ need not to commute with $L_1$.  Typically the generator of the perturbation is $L_1 = F_1 \cdot \nabla$ for states $x\in \bbR^d$.  We are thus interested in obtaining a useful relation for $\nabla e^{tL} A$.  In other words, we need to start from
\begin{equation}\label{rue}
\fbox {$ \displaystyle
R(\tau) = \int\,\rho(\id x) F_1(x)\cdot  \nabla e^{\tau L}A\,(x) 
$}
\end{equation}
Ruelle treated a formula with this form for deterministic dynamical systems~\cite{rue98,rue09}, recently applied
especially in studies of simple models for climate response ~\cite{abr07,abr08,luc11}.
Indeed (\ref{rue}) is especially suited in case the stationary probability law is {\it strange} in the sense that its lack of smoothness forbids further partial integration to go back to the Agarwal formula \eref{agar}.  In that same context however, it is
useful to further split \eref{rue} into two parts~\cite{rue09}, one describing fluctuations along the stable directions of motion and one parallel to unstable directions (those with positive Lyapunov exponents).  Informally you would expect that the vector $\nabla e^{\tau L}A\,(x)$ can be given, at least for large $\tau$, in its natural components $\exp[\tau \lambda_{\hat{e}}(x)]\,\hat{e}\cdot\nabla A(x)$ for local Lyapunov exponents $\lambda_{\hat{e}}$ corresponding to the various (stable and unstable) directions $\hat{e}$.  Such a decomposition is natural for stationary distributions $\rho$ which belong to the class of Sinai-Ruelle-Bowen (SRB) measures; these have a density in the unstable (expanding) direction (positive Lyapunov exponents) so that there a further {\it partial} integration remains possible towards an Agarwal formula \eref{agar}.  That is exactly what is suggested in the hybrid form of Section 2.3 of \cite{abr07}. 
See~\cite{lsy02,tas98} for an introduction on SRB-measures with their physical interpretation.

\subsubsection{Algorithm for deterministic and stochastic systems}
\label{sec:det-num}
The numerical evaluation of linear response for chaotic dynamics has recently been studied by various groups.  
In particular, Abramov and Majda have developed new computational approaches based on the theory of 
SRB-measures \cite{abr08}.   
We simplify here the presentation to introduce an algorithm that works also for stochastic systems. 
To exploit \eref{rue}, no knowledge of $\rho$ is required, 
but we need to get a hold on $\nabla_{\!x}e^{\tau L}A\,(x) =  \nabla_{\!x}
\langle A(x_\tau)\rangle_{x_0=x}$ where we have emphasized that the differentiation is on an evolved quantity
with respect to the initial condition $x$.  
To solve the problem that $\nabla$ (state derivative) and $L$ (generating the time-evolution 
$e^{tL}A(x)$) need not to commute, we unfold the formula a little further. 
A practical numerical tool for estimating $\nabla_{\!x}A(x_\tau)$ in deterministic dynamics 
(where $x_\tau$ is uniquely determined from $x_0=x$) was already
presented in \cite{eyi04}, but the following numerical method is more efficient for steady states.

Consider first an evolution in discrete time $n=0,1,\ldots$ (but with a parameter $\epsilon$ that will allow 
a continuous time limit $\epsilon \to 0$),
\begin{eqnarray}\label{dyns}
x_{n+1}= g_n(x_n) = x_n + &v_n(x_n)& ,\qquad x_0=x\\
\mbox{with} &v_n(x)& \!\!\!= \epsilon[F(x)  + \xi_n]\nonumber
\end{eqnarray}
for $\xi_n$ anything stationary in time, including possible ``noise'' depending on the time $n$, 
but that does not depend on the state $x$. This $(\xi_n)$ is considered frozen so that for its given realization
we take \eref{dyns} as a deterministic dynamics. The main point is that $\nabla_x\langle A(x_n)\rangle_x = \langle \nabla_x A(x_n)\rangle_x$ where the $\langle\cdot\rangle$ averages over the ``noise'' $(\xi_n)$. We then need to deal with $\nabla_{\!x} A(x_n)$ where $x_n$ depends on the initial state $x$ through \eref{dyns}.  By the chain rule, applied recursively,
\begin{eqnarray}
\nabla_x A(x_n) &=& (\nabla A)(x_n)\cdot\nabla_x \big(g_{n-1}\circ\ldots\circ g_0\big)\,(x)\nonumber\\
  &=& (\nabla A)(x_n)\cdot \bbG_{n-1}\,(x)\nonumber\\
  &=& (\nabla A)(x_n)\,(\nabla g_{n-1})(x_{n-1}) \cdot \nabla_x\big(g_{n-2}\circ\ldots\circ g_0\big)\,(x)\nonumber\\
  &=&(\nabla A)(x_n)\,(\nabla g_{n-1})(x_{n-1})\cdot \bbG_{n-2}(x)=\ldots \label{Gn}
\end{eqnarray}
where $\nabla A$ is a $1\times d$ row vector (easily computed at all times) and 
$\bbG_k(x) \equiv \nabla \big(g_k\circ g_{k-1}\circ\ldots\circ g_0\big)\,(x)$ is a $d\times d$ matrix obeying the
recursive relations
\numparts\label{eq:dro}
\label{recG}
\begin{eqnarray}
\bbG_n(x) &=& \nabla g_{n}(x_n) \cdot \bbG_{n-1}(x) \label{recGa}\\
\bbG_0(x) &=& \nabla g_{0}(x) \label{recGb}
\end{eqnarray}
\endnumparts
Note now that for each time $k$, the analytical form of the matrix $\nabla g_k = \bbI + \epsilon \nabla F$ 
does not depend on $k$ or on the ``noise'' $(\xi_n)$.  Hence,
\begin{eqnarray}
\label{nabla}
\nabla_x A(x_n) &=& (\nabla A)(x_n)\cdot
\left[\bbI + \epsilon(\nabla F)(x_{n-1})\right]\nonumber\\
&&\cdot\left[\bbI + \epsilon(\nabla F)(x_{n-2})\right]\cdots
\left[\bbI + \epsilon(\nabla F)(x)\right]
\end{eqnarray}
The continuous time version $\bbG_t(x)$ can be obtained in a suitable limit $n\to \infty$ for $\bbG_n(x)$ with
time $t=n\,\epsilon$ and time step $\epsilon= t/n$. In this limit (\ref{dyns}) returns the evolution 
$\dot{x}_t = F(x_t)$ for the deterministic case, while for stochastic equations an additional suitable rescaling 
to reproduce e.g. Gaussian noise with $\xi_t$ is needed. At  any rate, 
from $\bbI + \epsilon (\nabla F)(x) \simeq e^{\epsilon(\nabla F)(x)}$ we get formally 
\begin{equation}
\label{nablat}
\nabla_x A(x_\tau) = (\nabla A)(x_\tau)\cdot{\textbf{T}}\exp\left[\int_0^\tau ds (\nabla F)(x_s) \right]
\end{equation}
where each $x_s$ depends deterministically on $x$ for frozen $(\xi_r, 0\leq r < s)$, and
\textbf{T} indicates a time-ordered integral. The formula \eref{nablat} or its discretization \eref{nabla} is ready to be inserted into \eref{rue} where the $\langle\cdot\rangle$ will first average there over the possible 
``noise'' and then, with $\rho$, over the initial condition $x$. That provides the main algorithm for the analytic approach working on the
observable instead of on the probability distribution.

A similar expression can be found in the Appendix B of Ref.~\cite{eyi04} for the case of deterministic dynamics. 
In that work, however, the estimate of $ (\nabla A)(x_n)\bbG_t(x)$ was performed with an adjoint scheme,
by integrating  numerically backward in time the final value $(\nabla A)(x_n)$ with the equation 
 \begin{equation}
 \partial_t (\nabla A)(x_t) + (\nabla A)(x_t)\cdot (\nabla g)(x_t) = 0
 \end{equation}
(In~\cite{eyi04} they wrote the equation for the column vector rather than for the row vector $\nabla A$, and
a transpose of $(\nabla g)$ was used).
Although that scheme is equivalent to our (\ref{nabla}), it requires a CPU time $\caO(n^2)$ 
for a perturbation active during all $n$ iterations, as opposed to the $\caO(n)$ matrix multiplications (\ref{recG}) for estimating matrices $\bbG_k$ for $k\le n$. 
Our scheme is exploiting stationarity: the propagator from time $n-k$ to time $n$ is the same
as $\bbG_k$ from time $0$ to $k$.
In transient regimes we would lose this property and we would also need $\caO(n^2)$ operations.

A further study of such a numerical algorithm is contained in \cite{abr07,abr08}; in particular Appendix A of \cite{abr08} explains the derivation above (without the generalization to stochastic evolutions).  For questions of dealing in a similar context with the impact of stochastic perturbations, we refer to \cite{luc12}.

\subsubsection{Numerical illustration}
\begin{figure}[!bt]
\begin{center}
\includegraphics[angle=0,width=9cm]{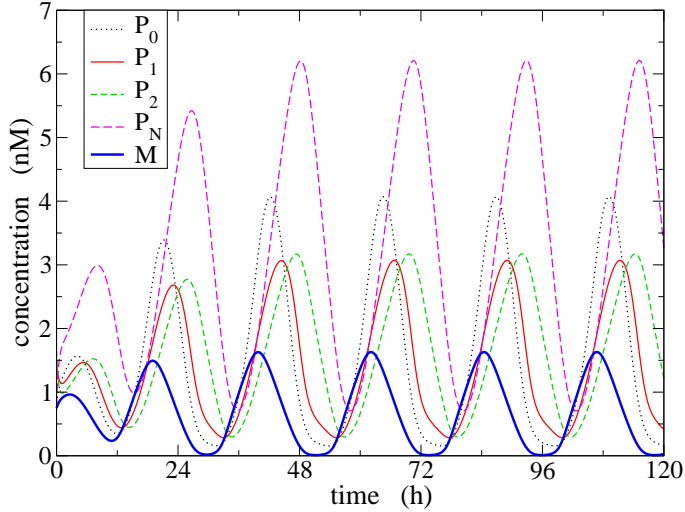}
\end{center}
\caption{Evolution of the concentration of mRNA and of the four proteins of the model from
random initial conditions, according to (\ref{eq:dro}).
\label{fig:cycle}}
\end{figure}

To illustrate the numerical scheme (\ref{Gn}) for estimating Ruelle's linear response formula
(\ref{rue}) in a simple context, 
we consider a set of equations introduced in biology to describe circadian cycles, that is the
periodicity of biorhythms, for Drosophila~\cite{gol95,oga06}.
The state space has $d=5$ dimensions, with states $x=(P_0$, $P_1$, $P_2$, $P_N$, $M)$. 
The dynamics couples the concentration $M$ of mRNA with
those of four types of proteins, written as $P_0$, $P_1$, $P_2$, and $P_N$ in Ref.~\cite{oga06}, 
where one can find the details of these equations.
Denoting
 \begin{equation}
\pfrac{P}{K} \equiv \frac{P}{P+K}
 \end{equation}
we have equations of motion $\dot x=F(x)$ of the form
\numparts\label{eq:dro}
\begin{eqnarray}
\frac{\id P_0}{\id t} &=& k_S M - \nu_1 \pfrac{P_0}{K_1}+ \nu_2 \pfrac{P_1}{K_2}\label{eqP0}\\
\frac{\id P_1}{\id t} &=& \nu_1 \pfrac{P_0}{K_1}- \nu_2 \pfrac{P_1}{K_2}- \nu_3 \pfrac{P_1}{K_3} + \nu_4 \pfrac{P_2}{K_4}\\
\frac{\id P_2}{\id t} &=& \nu_3 \pfrac{P_1}{K_3} - \nu_4 \pfrac{P_2}{K_4} +  \nu_d \pfrac{P_2}{K_d} -k_1P_2+k_2P_N\\
\frac{\id P_N}{\id t} &=& k_1P_2-k_2P_N\\
\frac{\id M}{\id t} &=& \nu_S \pfrac{K_I^n}{P_N^n} - \nu_m \pfrac{M}{K_m}
\end{eqnarray}
\endnumparts
with parameters as in previous papers,
\begin{eqnarray*}
&&\left(\nu_S = 0.5,\, \nu_m = 0.3,\, \nu_1 = 6,\, \nu_2=3,\, \nu_3=6,\, \nu_4=3\right)\, 
\textrm{nM\,h}^{-1}\\
&&\left(K_m=0.2,\, K_I=2,\, K_1=1.5,\, K_2=2,\, K_3=1.5,\, K_4=2\right)
\,\textrm{nM}\\
&&\left(k_S=2,\, k_1=2,\, k_2=1\right)\,\textrm{h}^{-1},\qquad n=4
\end{eqnarray*}
Integration of (\ref{eq:dro}) was performed 
with a simple Verlet scheme with a discrete time step $\epsilon=0.0025$.
Starting from random concentrations, the model reaches quickly a cyclic regime, as shown in
Fig.~\ref{fig:cycle}.

We check the response in the mRNA concentration $M$ to a change of the
rate $k_S\to k_S(1+h)$, starting from the steady state, i.e.,~a random phase of the cycle. 
The dynamical perturbation introduced by that change is 
$h\,F_1 = (h\, M\, k_S,0,0,0,0)$, since a change in $k_S$ affects only the evolution equation of variable $P_0$,
see (\ref{eqP0}).
The observable $A(x)=M$ has a gradient $(\nabla A)(x) = (0,0,0,0,1)$, which is coupled to the perturbation via
the ``propagator'' matrix $\bbG_n$.
With the setup of Sec.~\ref{sec:det-num}, we can estimate $\bbG_n$ by a sequence of matrix multiplications.
Given an evolution in discrete time $g(x) = x + F(x)\epsilon$,
we are interested in matrix $\nabla g(x) = \bbI + \nabla F(x) \epsilon$, or in coordinates
$(\nabla g)_{ij}(x) = \delta_{ij}+\,dF_i/dx_j\, \epsilon$. The rows of the matrix $(\nabla F)$ are

\numparts\label{nablaF}
\begin{eqnarray} 
&&\left(
-\nu_1\pfrac{P_0}{K_1}_{P_0} , 
\nu_2 \pfrac{P_1}{K_2}_{P_1}, 
0, 
0, 
k_S
\right)\\
&&\left(
\nu_1\pfrac{P_0}{K_1}_{P_0} , 
-\nu_2 \pfrac{P_1}{K_2}_{P_1}\!\!\!-\nu_3 \pfrac{P_1}{K_3}_{P_1},
\nu_4 \pfrac{P_2}{K_4}_{P_2} ,
0,
0
\right)\\
&&\left(
0, 
\nu_3 \pfrac{P_1}{K_3}_{P_1},
-\nu_4 \pfrac{P_2}{K_4}_{P_2}\!\!\!-\nu_d \pfrac{P_2}{K_d}_{P_2}\!\!\!-k_1,
k_2,
0
\right)\\ 
&&\left(
0, 
0, 
k_1, 
-k_2, 
0
\right)\\
&&\left(
0,
0, 
0, 
\nu_S\pfrac{K_I^n}{P_N^n}_{P_N},
-\nu_m\pfrac{M}{K_m}_{M}
\right)
 \end{eqnarray}
\endnumparts
where derivatives with respect to $P$ are denoted as
$
\pfrac{P}{K}_P = \frac{1}{P+K} - \frac{P}{(P+K)^2}
$
and similarly for $\pfrac{K_I^n}{P_N^n}_{P_N}$.
This matrix with $x_0=(P_0,P_1,P_2,P_N,M)_0$ yields 
$(\nabla g)(x_0) = \bbI + (\nabla F)(x_0)\epsilon $, coinciding
with $\bbG_0$. Iteratively, $\bbG_1 = (\nabla g)(x_1) \cdot \bbG_0$, and so on.

\begin{figure}[!bt]
\begin{center}
\includegraphics[angle=0,width=9cm]{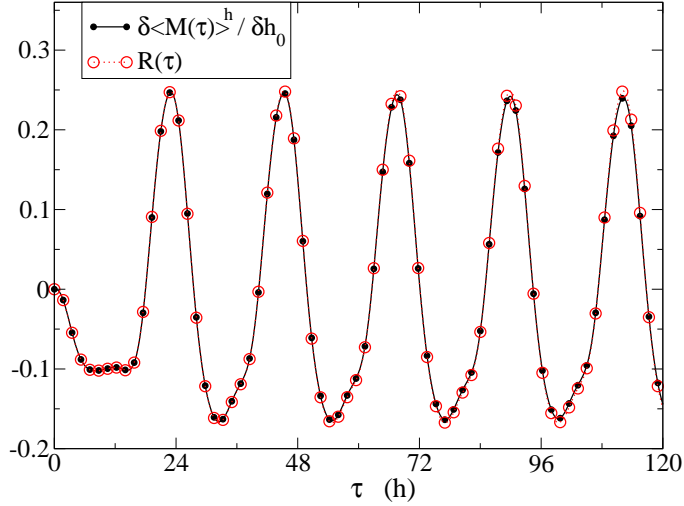}
\end{center}
\caption{Linear response function $R(\tau)$ of the mRNA density $M$ to an impulsive change in the parameter 
$k_S\to k_S(1+h_0)$ as a function of time and response $\delta\mean{M(\tau)}/\delta h_0$ computed
with $h_0=10^{-3}$. Units are nM vs. hours.
\label{fig:resp1}}
\end{figure}

By sampling many trajectories, in parallel for perturbed and unperturbed 
dynamics starting from an initial condition from the steady state, 
the estimate of $\delta M(\tau)/\delta h_0$ has been obtained by setting a small constant 
$h$ for $\tau>0$ and by evaluating the time derivative of
$(\mean{M(\tau)}^h -\mean{M(\tau)})/h $ (we used $h=10^{-3}$).
The response function instead has been calculated with (\ref{Gn}) and (\ref{rue}), and
well overlaps with the response, as shown in Fig.~\ref{fig:resp1}.
Asymptotic stability in this model~\cite{oga06} probably favors a good
performance of formula (\ref{rue}).
We computed both the perturbed dynamics and the unperturbed fluctuation formula 
just to show graphically that the results are the same, but of course there is no additional 
understanding of the process. 
Obviously, computing the response only with unperturbed simulations would be convenient 
as long as the convergence of the method is good. 
We postpone a more detailed study of the efficiency of this and other numerical
schemes to a future work.

\section{Probabilistic approach}\label{pro}
Linear response makes sense in general only within a statistical theory.  That is to say, sensitive dependence on initial and boundary conditions can create strong and relevant effects beyond linear order on the microscopic scale while macroscopic linearity remains valid.  For estimating the mobility we do not investigate the microscopic particle's individual motion at specific times and how it changes under an external field, but we ask for a spatio-temporal averaged current and that includes noise.  In fact, for stronger microscopic chaoticity we expect the statistical approach to be more relevant. While various physical variables can show chaotic behavior in their time-evolution, their spatial or temporal averages will typically have a much smoother behavior,  see also \cite{col12} and chapter 6.2 in \cite{dor99} answering the so called van Kampen objection.

\subsection{As noise gets important}\label{asn}
In coarsening dynamics of low temperature spin systems, or in spin glasses, a very long
 transient regime may exist towards equilibrium, i.e.,~the nonequilibrium is not imposed
by external gradients and the dynamic equations are actually equilibrium ones. 
It is in this context that an extensive response-literature has been produced 
during the last decades. We mention briefly some results from 2003,
whose focus was mostly to develop zero-field algorithms, in other words, 
exploiting fluctuation--response relations to estimate numerically responses from fluctuations 
in unperturbed dynamics.
Works of Chatelain~\cite{cha03}, Ricci-Tersenghi~\cite{ric03}, and Crisanti and Ritort~\cite{cri03}
 introduced new schemes for computing the response
of the system from correlations in the spins. They were followed by Diezeman~\cite{die05} and by 
Lippiello, Corberi, and Zannetti~\cite{lip05,and06,lip07}. 
The derivations of these results cannot be reduced in our brief scheme and 
various forms of response relations have been obtained.   
However, the one by Lippiello et al., a specific case of (\ref{st1}) below,
emerged as having clear physical significance beyond numerical usefulness, because it matched the form of 
an earlier study by Cugliandolo, Kurchan and Parisi~\cite{cug94}, who proposed a response relation 
for autocorrelations in Langevin equations. 
With the present notation (\ref{lan}), perturbation $h_s x$ and observable $A(x)=x$,
that would read
\begin{equation}\label{ckp}
\fbox {$ \displaystyle
R(t,s) = \frac 1 {2 D} \left\{ 
\left(\frac \id {\id s} - \frac \id {\id t}\right)\mean{x_s x_t} 
-\left( \mean{F(x_t) x_s} -  \mean{F(x_s) x_t}\right) 
\right\}
$}
\end{equation}
showing an ``asymmetry'' term in addition to the form of the Kubo formula.
The derivation of (\ref{ckp}) is based on the equality $\mean{x_t\xi_t}=2 D \beta R(t,s)$, 
hence it is centered on the presence of noise $\xi_t$. This was new, certainly with respect to derivations
presented in previous sections where eventually noise was just an aspect of the dynamics, 
not fundamental for the derivations.
An equation similar to (\ref{ckp}) appears in a more recent study for nonequilibrium 
steady states~\cite{har06}, based on a path-space formulation by Harada and Sasa~\cite{har05}, see their Appendix B.  
The Harada-Sasa approach has really pioneered the path-space approach of the next subsection.
The consistency between the approaches of Refs.~\cite{cug94,lip05,har05} indeed shows that these belong
to a general framework with new significant physical content.
In the following section we discuss the path-space formulation embracing these results and 
we mention the physical interpretation that goes beyond merely dissipative aspects.  Noise becomes visible then as dynamical activity, ruling the time-symmetric fluctuations.

\subsection{Path space approach}\label{path}
The origin of dynamical ensembles is the projection of a microscopic Hamiltonian dynamics on the dynamics of reduced variables.  That Mori-Zwanzig projection~\cite{zwa61,mor65} originates from making a physical partition of the phase space, depending on the physical situation at hand, in which each microstate $X$ is mapped (many--to--one) to a reduced state $x(X)$. That induces noise in the reduced dynamics, for example on the mesoscopic level of description.  It is then natural to consider dynamical ensembles, i.e., probability distributions on path space where a path refers to a trajectory on the mesolevel. Such was already the approach of Onsager and Machlup starting dynamical fluctuation theory in \cite{ons53}.  These trajectories, under certain limiting conditions (e.g. via a weak coupling limit or via adiabatic elimination), can be described via first-order equations, in which case we meet the Markov processes of the previous Section \ref{per}.  Dynamical ensembles can however also describe non-Markovian processes describing important memory effects, see e.g.~\cite{boh12} for the application of a response relation in a visco-elastic medium.

Paths are trajectories $\omega = (x_s, 0\leq s\leq t)$ in state space, say looking at the states $x_s$ in the time-interval $[0,t]$.  For the sake of simplicity we characterize the unperturbed (perturbed) process by the probability weight $P(\omega)$ [$P^h(\omega)$] with respect to some reference $\id \omega$. The mathematical idea is to turn perturbed expectations 
$\mean{A(t)}^h$ into unperturbed ones $\mean{A(t) P^h(\omega)/P(\omega)}$.
Defining the relative action $\caU(\omega) = \log P(\omega)/P^h(\omega)$ and splitting 
$\caU(\omega) = [\caT(\omega)-\caS(\omega)]/2$ 
into a time-symmetric $\caT(\omega)$ and a time-antisymmetric $\caS(\omega)$, we get
\numparts\label{dynen}
\begin{eqnarray}
\langle A(t)\rangle^h - \langle A(t) \rangle &=& 
\int \id \omega \,P(\omega) A(x_t) \big( e^{-\caU(\omega)} - 1\big) \\
&=& \mean{A(t) \,\big( e^{-\caU(\omega)} - 1\big)}\\
&=&  \frac 1 2 \mean{A(t)\caS_1(\omega)} -  \frac 1 2 \mean{A(t)\caT_1(\omega)}+\caO(h^2)\label{ST}
\end{eqnarray} 
\endnumparts
where $\caT_1$ and $\caS_1$ are the linear contributions in $h$ of $\caT$ and $\caS$, respectively,
around $h=0$ (where $\caU=0$).  This formula is general but the point is  that in physical systems with local detailed balance \cite{har05}  we have a good physical understanding of the path functions $\caT$ and $\caS$~\cite{bai10,bai09,bai09b,mae10}.  Let us first look under global detailed balance.
 We already see here that since the observable $A(t)-A(0)$ is time-antisymmetric 
(ignoring momenta), under detailed balance
\begin{eqnarray}\label{adynen}
\langle A(t)\rangle^h - \langle A(t) \rangle &=& \langle A(t) - A(0)\rangle^h \nonumber\\
&=& \frac 1 2 \mean{[A(t)-A(0)]\caS_1(\omega)} \nonumber\\ 
&=& \mean{A(t)\,\caS_1(\omega)} 
\end{eqnarray}
because $\mean{[A(t)-A(0)])\caT_1(\omega)} = 0$ and $\mean{A(t)\,\caS_1(\omega)} = -\mean{A(0)\,\caS_1(\omega)}$ by time-reversal symmetry of the reference equilibrium process.
The result \eref{adynen} should equal the Kubo formula \eref{kubo}, and indeed it does as will become clear.
When the perturbation is from a potential $-h_s V(x_s)$,
it is well established that the time-antisymmetric $\caS(\omega)=\caS_1(\omega)$ must be the path-dependent
entropy flux into the environment as caused by the perturbation potential $V$~\cite{mae03,maes03}.
If the environment is at uniform temperature, for small constant $h$ for times $t>0$ the change in entropy is
$\caS(\omega)=\beta h [V(t)-V(0)]$, namely dissipated energy $h[V(t)-V(0)]$ divided by temperature. 
The first term on the right-hand side of \eref{ST}  has thus the same form one finds in equilibrium \eref{adynen}, but in general the unperturbed process is in steady nonequilibrium and
 $\caS(\omega)=\beta h [V(t)-V(0)]$ is an excess of 
entropy production with respect to that already generated by the nonequilibrium dynamics. 
Hence, the derivative $\delta\mean{A(t)\caS(\omega)}/\delta h_s=\beta\frac{\id}{\id s}\mean{V(s) A(t)}$ 
also mimics the equilibrium Kubo formula.

We now turn to the second term on the right-hand side of \eref{ST}. 
Eq.~(\ref{ST}) must be true also for a constant $A$, for which the response is zero.
We thus get 
 \begin{equation}
\beta \frac \id {\id s}\mean{V(s)} = \frac \delta {\delta{h_s}} \mean{\caT_1(\omega)} =  \mean{\frac \delta {\delta{h_s}}\caT_1(\omega)}
 \end{equation}
If $\frac \delta {\delta{h_s}}\caT_1(\omega)$ equals a state function $B(s)$, 
then, since  $\frac \id {\id s}\mean{V(s)} =  \mean{LV(s)}$,
one deduces that $B(s)=\beta LV(s)$, arriving at
\begin{equation}\label{st1}
\fbox {$ \displaystyle
R(t,s) = \frac \beta 2 \frac \id {\id s} \mean{V(s)A(t)}
- \frac \beta 2 \mean{L V(s)A(t)}
$}
\end{equation}
For diffusion processes \eref{lan} with generator \eref{gene} and temperature $D=1/\beta$ 
we see that indeed $LV(x) = (F\cdot \nabla V)(x) + D \Delta V(x)$ is a state function. 
The same is true for Markov jump processes (for specific derivations of \eref{st1} 
using stochastic calculus, see~\cite{bai09b}).
The function $LV$ quantifies the time-symmetric volatility of $V$ under the unperturbed dynamics, also named
``frenesy''. It is related to the ``dynamical activity'' in discrete systems, 
where it quantifies the change in escape rates; $\cal T(\omega)$ is the excess in dynamical activity 
over $[0,t]$. 
We refer to ~\cite{bai09,bai09b,mae10} for more details and examples.  
Thus, there is a new contribution next to the first term in (\ref{st1}) 
(taking one half of the Kubo formula \eref{kubo} and referring as usual to dissipation). In particular, its form now clearly deviates from response formul{\ae} such as \eref{sei}.  One then wonders in what sense the decomposition in formul{\ae} like \eref{st1} is intrinsically  natural.  We think it is, as that splitting makes the first term time-antisymmetric and the second term symmetric in time $s$, which corresponds to the dissipative and the reactive part of the susceptibility, respectively.

Note that in (\ref{st1}) one computes averages without needing an 
explicit knowledge on the stationary probability. On the other
hand some knowledge is required on the dynamics sitting in the generator $L$. 
 It is still not clear how much kinetic information is truly needed and how practical that gets.  
We summarize the general physical idea in the formula
\begin{equation}\label{dec2}
\langle A(t)\rangle^h - \langle A(t)\rangle = 
  \frac 1{2}\langle \mbox{Entr}^{[0,t]}(\omega)\,A(t)\rangle
- \frac 1{2}\langle \mbox{Esc}^{[0,t]}(\omega)\,A(t)\rangle
\end{equation}
where Entr$^{[0,t]}(\omega)$ is the excess in entropy flux over the time period $[0,t]$ and 
Esc$^{[0,t]}(\omega)$ is the excess in dynamical activity due to the perturbation. 
The physical challenge is to learn to guess or to find that Esc$^{[0,\tau]}(\omega)$ from partial information on the dynamics.  For the overdamped dynamics of e.g. \eref{st1} it means to consider the expected rate of change of the perturbing potential $V$ under the original dynamics.  The response relation out of equilibrium is no longer a fluctuation-dissipation relation but a fluctuation--dissipation-activation relation.  In fact, the formula can now be turned around and from measuring violations of the fluctuation--dissipation relation one obtains information about the active forces~\cite{boh12}.

\subsection{Inertial case}
We open a separate subsection on the inertial case of Langevin dynamics because its linear response is much less discussed in the literature despite the obvious interest for example for models of heat conduction.  The main point however is that the ideas summarized under \eref{dynen} and \eref{dec2} remain unchanged.

For states $(q,p) =
(q^1,q^2,\ldots,q^n;$ $p^1,p^2,\ldots,p^n)\in \bbR^{2n}$ of positions
and momenta we attach standard white noise $\xi^i_t$ to each $1\leq i\leq n$, with constant strength $D^i$ and a
friction coefficient $\gamma^i$ to model heat baths at
temperature $D^i/\gamma^i = T^i$:
\begin{eqnarray}
\dot{q}^i &=& p^i \nonumber\\
\dot{p}^i &=&
 F^i(q)  -\gamma^i p^i  +
  h_t \frac{\partial V}{\partial q^i}  + \sqrt{2D^i}\,\xi^i_t
\label{ud}
\end{eqnarray}
The forces $F^i$ can contain a nonconservative part but are confining when we want a stationary regime where the particles typically reside in a bounded region.  We already inserted the perturbation $V(q)$ with small
 time-dependent amplitude $h_s$ for $ s\geq 0$.  The linear response is given by formula (17) 
in \cite{bai10}:
 \begin{eqnarray}\label{genfor}
 R(t,s) =& \sum_{i}\frac 1{2T^i}&\!\!\!\!\!\!\!\!\!
\mean{\frac{\partial V}{\partial q^i}(q_s) \,p^i_s\,A(t)} \nonumber\\ 
&-\sum_i\frac 1{2D^i} \Bigg\{&
\mean{\frac{\partial V}{\partial q^i}(q_s)\,F^i(q_s)\,A(t)}\nonumber\\
 &&
 - \frac{\id}{\id s} \mean{\frac{\partial V}{\partial q^i}(q_s)\, p^i_s\,A(t)}\nonumber\\
 &&
 + \sum_j \mean{\frac{\partial^2 V}{\partial q^j\partial q^i}(q_s)\,p^j_s\,p^i_s\,A(t)}
\Bigg\}
 \end{eqnarray}
The first sum again corresponds to the dissipative part from the entropy fluxes in the reservoirs at temperatures $T^i$.  The remaining sums give the frenetic contribution. As not recognized yet, formula \eref{genfor} can still be rewritten in a similar way as done in \eref{st1}.  Supposing $D^i=\gamma/\beta$ we must replace there $V\rightarrow \dot{V}/\gamma =p\cdot \nabla_q V/\gamma$,
\begin{equation}
\label{st2}
\fbox {$ \displaystyle
R(t,s) = \frac {\beta}{2\gamma} \frac \id {\id s} \mean{p_s\cdot \nabla_q V(s)A(t)}
- \frac {\beta}{2\gamma} \mean{L(p_s\cdot \nabla_q V)(s)\,A(t)}
$}
\end{equation}
in which the generator $L$ for \eref{ud} now reads
 \begin{equation}
Lf(q,p) =
p\cdot\nabla_q f + (F-\gamma\,p)\cdot\nabla_p f + 
D \Delta_p f
 \end{equation} 
and is the underdamped version of \eref{gene}. Note however that the dissipative last term $D \Delta_p f$ does not contribute in \eref{st2} and that these formul{\ae} do not work in case  $D^i=0$ for an $i$ with $\partial V/\partial q^i \neq 0$.  In that case the best alternative is probably to apply the algorithm of Section \ref{sec:det}.

An important application of \eref{genfor} is to the modification of the Sutherland-Einstein formula relating transport coefficients such as mobility with fluctuation quantities such as the diffusion constant~\cite{bai11}.

\section{Further remarks and conclusions}
\label{rema}

\subsection{Nonlinear responses}
There is also a growing  number of works on higher-order terms around equilibrium.  In fact part of the book by Evans and Morriss is devoted to that~\cite{eva90}.  Other references include \cite{and07}, Section 10 in \cite{maes03} or the more recent \cite{mar08,mal11,lip08,vil09}. One typical start is
the fluctuation symmetry in the distribution of the entropy flux, transient from the reference equilibrium system as also explained in \cite{eva08} for thermostated dynamics.

 As was emphasized in \cite{col11} the main point is probably not to be able to write formal expansions and formul{\ae}, but to find useful structures and unifying interpretations.  We will not deal with that here, except for mentioning one particular general relation between second order and first order terms that has been largely unnoticed \cite{sai12}, and an instance of which has appeared as identity (25) in \cite{col11}.
\begin{equation}\label{2r}
 \left.\frac{\partial^2}{\partial h^2}\langle A(t)\rangle^h\right|_{h=0}
= \beta\left.\frac{\partial}{\partial h} \langle [A(t)-\bar{A}(0)][V(t)-V(0)]\rangle^h\right|_{h=0}
\end{equation}
The $\bar{A}$ equals $A$ up to flipping the sign of the momenta.
All terms  are explicitly expressed as correlation functions in the equilibrium reference process (indicated with the superscript $\langle\cdot\rangle^0$). The derivation of \eref{2r} uses linear response around nonequilibrium for non-state functions (i.e., the correlation function in the right-hand side).

\subsection{Non-state functions}
In case of observables that depend explicitly on time or are functions of the trajectory 
over several times (such as products $A_1(t_1)A_2(t_2)\ldots A_n(t_n)$), some of the formul{\ae} must change.
The basic techniques remain however in place.  
A  typical application is how the heat depends on a change in parameters, e.g., for estimating nonequilibrium heat 
capacities~\cite{bok11}.  Heat is not a state function but varies with the trajectory and the applied protocol.  

\subsection{Note on effective temperature}
A traditional approach to violation of the equilibrium fluctuation-dissipation theorem is to imagine an effective temperature in the otherwise unchanged Kubo formula~\cite{cug11}.
That idea has had most success with mean field type systems, but it has remained more unclear how the effective temperature can provide a consistent and general tool for realistic systems.   Still today it serves as a paradigm for interpreting experimental results, see e.g. \cite{ben11}.  One possible approach for the future would be to associate an effective kinetic temperature to the ratio between the frenetic and the entropic contribution in \eref{dec2}, see Appendix A in \cite{bai09b}.

\subsection{Transient case}
Lots of attention have been devoted to the linear response behavior for relaxational processes. 
We mentioned some of that in Section~\ref{asn}, but there is no way to be complete.  For example, the interest in ageing and glassy dynamics has much stimulated the search of modified fluctuation-dissipation relations~\cite{cal05}.  Here we emphasize that the analytic approach and in particular the methods of Section \ref{apr} loose their simple structure when the unperturbed reference is time-dependent (and not stationary as was assumed).  On the other hand, formul{\ae} like \eref{st1} are unchanged when one is not starting at time zero from the stationary distribution, but one needs to take the average in the unperturbed transient regime.  Also causality is automatically verified there as the second term in \eref{st1} equals the first term when $s > t$.  This unification in expressions for the transient and the steady regime is only natural as even the stationary regime is physically and ultimately but a very long transient. 

\subsection{General properties of linear response}
Some formal properties of linear response go basically unchanged from equilibrium to nonequilibrium contexts.  For example, sum rules and Kramers-Kronig relations only depend on the presence of an underlying Hamiltonian dynamics for the total system plus environment or on causality.  Of course, specific expressions can differ but there is physically nothing new, see also \cite{shi10,shi10b}.  Nevertheless, there do exist essential differences.  We have already alluded to the fact that the name ``fluctuation--dissipation theorem'' is no longer so appropriate because of the importance and complimentary character of changes in the dynamical activity (and not only in the dissipation).  In  fact, the word ``fluctuation'' also becomes less correct as the time-correlations in the expressions for 
$R(t,s)$  no longer express a symmetrized time-correlation. 
 To make that point, let us evaluate the equilibrium Kubo formula 
\eref{kubo}) when the perturbing potential $V$ equals the observable $A$:
\begin{equation}
\langle V(t)\rangle^h - \langle V\rangle = -  \langle V(0)[V(t)- V(0)]\rangle = 
\frac 1{2}\langle [V(t)- V(0)]^2\rangle  \geq 0
\end{equation}
which is the variance (or the ``diffusion'') of the displacement $V(x_t)-V(x_0)$. That is sometimes called a generalized Einstein relation.  In fact all linear transport coefficients for equilibrium can be expressed like that, as was understood already in 1960 by Helfand~\cite{hel60}. For nonequilibrium linear response, that relation including its positivity gets violated, see also \cite{bai11,mae11}.  In the words of the previous subsection, negative effective temperatures become possible in nonequilibrium.  The origin lies in the frenetic contribution, e.g. the second term on the right-hand side of \eref{st1}, which can overrun the first dissipative term. 

 \subsection{Outlook}
Response theory is primarily about predicting the reaction of a system in terms of its unperturbed behavior.  As we mostly have in mind response in time, that involves temporal correlations.  Therefore, {\it dynamical} fluctuation theory can be expected to be most prominent.  
(See \cite{mae07} for the distinction between static and dynamic fluctuation theory: statics looks at deviations around the law of large numbers at single times, for example for the average over many copies of the system, while dynamical fluctuations are around the law of large times, for deviations around time-averages.)   Around equilibrium,  dynamical fluctuations  are governed, just as static fluctuations, by the entropy and dissipation functions.  That further enables connecting different response coefficients, such as in Onsager reciprocity or via Maxwell relations.
Nonequilibrium makes a more drastic difference between static and dynamical fluctuations.  In particular, no useful connections between different types of responses have been discovered for nonequilibrium processes.  We think of the analogue of relations between compressibilities, heat capacities and conductivities.  In our opinion, that challenge will require finding experimental access to quantities like the dynamical activity which is complementing entropic characterizations in the description of nonequilibrium processes.

\subsection{Conclusions}

We have presented a concise guide (with plenty of references) to the multi-faceted world of linear response for systems out of equilibrium. From there we have discovered similarities but also some missing pieces that otherwise would not be evident. One such piece was the extension to stochastic systems of Ruelle's formulation, which we have introduced, together with an efficient algorithm. That also allows for numerical calculations where an explicit knowledge of the density of states is not required, as opposed to other (analytic) formulations that we have described. However, the resulting linear response formula contains a correlation function whose physical meaning is not very clear.

 In contrast, a rich physical picture emerges from a probabilistic approach based on path space weights, where the stationary distribution is also not needed. Besides the quite different mathematical apparatus compared to the more standard analytical approach, the probabilistic way indeed emerges as the one that currently offers more relations with dynamical fluctuation theory:  one has to study also how the system correlates with the activity of the perturbing potential, a time-symmetric quantity complementary to the time-antisymmetric fluxes of entropy. The combination fluctuation-dissipation therefore does not suiffice to characterize the linear response of nonequilibrium systems.

\paragraph{Acknowledgments}
M.B. thanks M. Colangeli, A. Vulpiani and J. Wouters for useful discussions.

\section*{References}

\bibliographystyle{./iopart-num2}


\providecommand{\newblock}{}

\end{document}